\documentstyle[epsfig,psfig,12pt,emulateapj]{article} 

\lefthead{NAIK ET AL.}
\righthead{RAPID STATE TRANSITIONS IN GRS~1915$+$105}

\slugcomment{To appear in The Astrophysical Journal}

\begin{document}

\title{DETECTION OF A SERIES OF X-RAY DIPS ASSOCIATED WITH A RADIO FLARE 
IN GRS~1915$+$105 }

\author{ 
	S.~Naik\altaffilmark{1},
	P.~C.~Agrawal\altaffilmark{1},
	A.~R.~Rao\altaffilmark{1},
 	B.~Paul\altaffilmark{1,2},
	S.~Seetha\altaffilmark{3} and
	K.~Kasturirangan\altaffilmark{3}}
\altaffiltext{1}
{Tata Institute of Fundamental Research,
Homi Bhabha Road, Mumbai 400 005, India}
\altaffiltext{2}
{Institute of Space and Astronautical Science
3-1-1 Yoshinodai, Sagamihara, \\
$~~~~~~~~~$Kanagawa 229-8510, Japan}
\altaffiltext{3}
{ISRO  Satellite Center, Airport Road, Vimanapura P.O., Bangalore - 560017, India\\
$~~~~~~~~~~~~~$sachi@tifr.res.in, pagrawal@tifr.res.in, arrao@tifr.res.in, bpaul@tifr.res.in}

\begin{abstract}
We report the detection of a series of X-ray dips in the Galactic black 
hole candidate GRS~1915$+$105 during 1999 June 6$-$17 from observations 
carried out with the 
Pointed Proportional Counters of the Indian X-ray Astronomy Experiment 
on board the Indian satellite IRS-P3. The 
observations were made after the source made a transition from a 
steady low-hard state to a chaotic state which occuered within
a few hours. Dips of about 20$-$160 seconds duration
are observed on most of the days. The X-ray emission outside the dips
shows a QPO at $\sim$4 Hz which has characteristics 
 similar to the ubiquitous 0.5 $-$ 10 Hz QPO
seen during the low-hard state of the source.
During the onset of  dips this QPO is absent and also the energy
spectrum is soft and the variability is low 
compared to the non-dip periods. These features 
gradually re-appear as the dip recovers. The onset of the 
occurrence of a large number of such dips followed  
the start of a huge radio flare of strength 0.48 Jy (at 
2.25 GHz).
We interpret these dips as the cause for mass ejection due to the 
evacuation of matter from an accretion disk around the black hole.
We propose that a super-position of a large number of such
dip events  produces a    huge radio jet in GRS 1915$+$105. 
 
\end{abstract}

\section{INTRODUCTION}

The X-ray transient source GRS~1915+105 was discovered by Castro-Tirado, 
Brandt, and Lund (1992) with the WATCH all sky X-ray monitor on-board the 
Granat satellite. The source has been exhibiting a wide variety of temporal 
variability in its X-ray and radio emission. GRS~1915$+$105, the first 
Galactic superluminal radio source, has characteristics of a micro-quasar
and is located at a distance of about 
12.5 $\pm$ 1.5 kpc (Mirabel \& Rodriguez 1994).

Besides the chaotic variability, narrow quasi-periodic oscillations (QPO) 
at centroid frequency  in the 
range of 0.001 $-$ 10 Hz were discovered in the X-ray emission  
from the source using the Indian X-ray Astronomy Experiment (IXAE) (Agrawal et 
al. 1996) and the RXTE (Morgan \& Remillard 1996). A QPO at a
centroid frequency of 67 Hz which does not change with time, was also 
detected in this source in the RXTE observations. It has been
 suggested that it is
related with the innermost stable orbit in the accretion disk of 
the source. Chen, Swank, \& Taam (1997) found that the
intensity dependant narrow QPOs are a  characteristic feature of the hard 
branch and it is absent in the soft branch which corresponds to the very 
high state similar to those of other black hole candidates.  
Trudolyubov, Churazov, \& Gilfanov (1999) studied the  1996/1997 
low luminosity state and transitions of states using the RXTE data. They
found a strong correlation between the QPO centroid frequency and 
spectral and timing parameters similar to the one detected in other
Galactic black hole candidates in the intermediate state. 
Muno, Morgan, \& Remillard (1999) sampled the RXTE data over a wide range 
of properties and found that the 0.5 $-$ 10 Hz QPOs are correlated with the 
temperature  of the accretion disk. From the studied behavior of the source 
they distinguish two different states of the source: 
the spectrally hard state, dominated by a power law component when the QPOs
are present and the soft state, dominated by thermal emission when the 
QPOs are absent.

Paul et al. (1998) detected quasi periodic bursts with period of about
45 s using the IXAE data obtained in 1997 June $-$ August 
and interpreted the slow rise and  fast decay of
bursts as evidence for matter
disappearing into the event horizon of the black hole. Yadav et al. (1999)
made a systematic analysis of these bursts and classified them into regular,
irregular and quasi-regular based on the burst duration and recurrence time.
They found the bursts to recur at a mean time of 20 $-$ 150 s
and they suggested that the irregular long duration bursts 
(recurrence time $\sim$ 120 s) during which
the spectrum becomes harder and harder as the burst progresses and
becomes hardest at the end of the decay,
are characteristic of the change of state of the source.
They calculated the rise time and the decay time for these
bursts to be of the order of a few seconds ($\sim$ 10s). 
The back and forth switching of the state of the source from a low-hard to 
a high-soft state within a time scale of a few seconds 
is explained by invoking the appearance and disappearance of the 
advective disk over its viscous time scale.
These irregular bursts were also quasi-simultaneously
observed by Belloni et al. (1997) who interpreted them as
repeated filling and evacuation of inner accretion disc.

Simultaneous X-ray and infrared observations of the source established
a close link between the non-thermal infrared emission and the 
X-ray emission from the accretion disk (Eikenberry et al. 1998). They
found that the X-ray properties showed a drastic change coincident with 
the occurance of the IR and radio flares. Similar episodes of X-ray and radio
flares were detected by Feroci et al. (1999) using the BeppoSAX satellite.
From the spectral analysis of the X-ray data, they found evidence for a 
temporary disappearance and subsequent restoring of the inner accretion 
disk during the flare. 
Such simultaneous multi-wavelength observations establish the disk-jet 
connection in GRS 1915$+$105. These observations, however, pertain to
jet emission which can be termed as ``baby-jets'' (Eikenberry et al. 1998)
from consideration of energy. On the other hand, the accretion disk
phenomena giving rise to superluminal jets are not very clearly
established. Harmon et al (1997)
found an anti-correlation between the decrease in the hard X-ray flux
(obtained from BATSE with a time resolution of $\sim$1 day) 
from the accretion disk and the subsequent jet production.
Fender et al. (1999) worked back the time of occurrence of super-luminally
moving jets and found rapid (20 $-$ 30 minute) radio oscillations
during the beginning of such jets. Though they attempted to associate
these oscillations with the X-ray oscillations detected by Belloni et al.
(1997), there are no simultaneous high time resolution observations
in the radio and X-ray wave-bands during huge radio flares responsible
for super-luminal jets.
Though there are some indication of association between radio and X-ray
emission based on low time resolution observations (RXTE ASM and GBI),
detailed quantitative association between the two is not very conclusive.
Hence we can conclude that the disk-jet connection for jets with
superluminal motion is indirect at best.

In this paper we report the detection of multiple 
X-ray dips interpreted as ``disk-evacuation''
events which are similar in nature to the ``baby-jet'' X-ray events
(Mirabel et al. 1998; Eikenberry et al. 1998).
These events started during the transition of the source state
from a steady low-hard state with the X-ray flux at 0.62 Crab at 1999 June 07 
17.79 UT to a chaotic state with a flux of 1.46 Crab at 19.22 UT. 
A simultaneous increase in radio flux was seen 
at 2.25 GHz from 0.029 Jy at 1999 June 07 10.94 UT to 0.478 Jy on 
1999 June 08 05.54 UT (from the public domain data from the 
NSF-NRAO-NASA Green Bank Interferometer). The presence of multiple dips 
in the X-ray light curve after the transition of the state when a radio
flare also occurred, suggest the presence of  ``disk-evacuation'' process in the
inner accretion disk of GRS 1915$+$105.   

\section{INSTRUMENT AND OBSERVATIONS }

The X-ray observations of  
GRS 1915$+$105 were carried out with the Pointed Proportional Counters (PPCs)
of the Indian X-ray 
Astronomy Experiment(IXAE) on board the Indian satellite IRS-P3.
The IXAE includes three co-aligned and identical, 
multi-wire, multi-layer proportional counters, filled with a gas
mixture of 90\% Argon and 10\% Methane at a pressure of 800 torr, 
with a total effective area of 1200 cm$^{2}$, covering 2 to 18 keV 
energy range with an average detection efficiency of about 60\% at 6 keV.  
The accepted X-ray events are stored in counters in 2$-$6 keV and 
2$-$18 keV band for the top layer, 2$-$18 keV band for the middle layer, 
${>}$ 18 keV (ULD counts) for all layers and ${>}$ 2 keV counts from the 
veto layers. For a detailed description of the PPCs refer to Agrawal (1998) 
and Rao et al. (1998).

The IRS-P3 satellite is in a circular orbit at an altitude of 830 km and 
inclination of 98$^{\circ}$. Pointing towards any particular source is 
done by using a star tracker with an accuracy of \(\leq\) 0$^{\circ}$.1. 
The useful observation time is limited to 5 of the 14 orbits outside 
the South Atlantic Anamoly (SAA) region to the latitude band of 
30$^{\circ}$S to 50$^{\circ}$N to avoid the charge particle 
background. The high voltage to the 
detectors is reduced when the satellite enters the SAA region 
and the data acquisition is stopped.

Background observations were made at the end of the source observation
by pointing the PPCs to a source-free region in the sky, close to the 
target source as the simultaneous background observation is not possible
because of the co-alignment of the PPCs.
The source GRS 1915$+$105 was observed from 1999 June 6 $-$ 15 with 
1 s time integration mode and June 16 $-$ 17 in 0.1 s time bin, for
useful period of 37,740 seconds. The log of the observations is 
given in Table 1.  

\section{ANALYSIS AND RESULTS}

The data, corrected for background and pointing offset, for PPC-1 and PPC-3, 
were added to construct the X-ray light curve for  GRS 1915$+$105. 
The dead time correction is not done as it is less than 1\% for each PPCs 
as the event processing time for each PPCs is about 20 $\mu$s. The X-ray
light curve for the source in the energy range of 2 $-$ 18 keV for all the
observations with the PPCs averaged over each orbit, is shown in Figure 1 (top 
panel). Also seen in the figure are 1.3 $-$ 12.2 keV energy range light 
curve obtained with the RXTE-ASM (middle panel) and  the radio 
flux at the frequency 2.25 GHz (bottom panel). 
The radio observations were taken from the 
public domain data from the NSF-NRAO-NASA Green Bank Interferometer.

  As can be deduced from the low resolution ASM and GBI data,
the source was in a steady low-hard state from June 2 to June
7 (MJD 51331 to 51336). It made a transition to a `chaotic state'
in a very short time which is very different from the slow
transition (about three  months) of the source in 1997
May to July (Trudolyubov, Churazov, \& Gilfanov 1999).
The source count rate increased from
about 53 ASM counts s$^{-1}$  on MJD 51336.5 (June 7 12.00 UT)
to 109.4 ASM counts s$^{-1}$ on MJD 51336.8 (June 7 19.2 UT).
The onset of X-ray activity is further constrained by the PPC
observations on June 7 17.50 UT, when the source showed a
steady behaviour. Hence we can conclude that there was an abrupt change
in the X-ray emission characteristics in less than 1.7 hours.
From the radio light curve of the source, a sharp
peak is clearly visible on MJD 51337 (June 8),
on the same day when the source showed the X-ray transition as seen from the
RXTE/ASM light curve.
The onset of the radio flaring event is between June 7 10.94 UT and
June 8 05.54 UT.
After the radio peak, on MJD 51337, the flux decayed slowly over next 5 days.
The exponential decay time of the radio flare is estimated to be 2.8 days.
Though the low time resolution X-ray light curve appears like a flaring event,
a closer examination of the high resolution data
reveals that variations of much shorter duration are also present as indicated
by a factor of 4 $-$ 5 variation over a few days within a span of
a few hundred seconds in a series of dips.

The light curves for some of the individual observations of the source 
with the PPCs are shown in Figure 2 and 3. No bursts or dips are seen in 
the X-ray light curves of 1999 June 06 and 07 (MJD 51335 and 51336) when 
the source was in a low-hard state. From 1999 June 08 onwards, various 
types of dips of duration in the range of 20 to 160 seconds are seen in
most of the observations as shown in Figure 2 and 3. During the dips, 
the X-ray flux decreases by a factor of about 3 within about 5 seconds, 
remains low for 20 to 160 s and then slowly recovers
to the maximum. The short term variability in the
X-ray light curves decreases during the dip. The duration of the 
dips is not constant in all the observations. In some of
the observations, short period dips of 20 $-$ 60 seconds duration are observed 
after the occurance of long dip with duration of more than 100 s. Towards 
the end of the observation, long duration dips lasting for more than 100 s 
are not seen but larger numbers of shorter duration dips are observed. 
A summary of the duration of dips observed in the
light curves of the source is given in Table 1.  

The first three panels of Figure 4 shows one of the dips observed 
during the PPC observations in
2 $-$ 18 keV, 2 $-$ 6 keV and 6 $-$ 18 keV energy ranges, co-added by 
matching the falling part of the dips. In the figure, 
the sharp decrease in the X-ray flux to almost one third of its original value
followed by a slow recovery is clearly seen.  We have calculated
the rise time and decay time for the dips by fitting exponential to the
light curve. The values are found to be about and 110 seconds and 7 seconds
respectively. The hardness
ratio (counts in 6 $-$ 18 keV energy range/counts in 2 $-$ 6 keV energy range)
for the same observation is shown in the fourth panel of the figure. 
It is seen from the figure that during the dips the spectrum 
(hardness ratio) of the source is soft. The bottom panel in Figure 4
shows the rms variation in the source calculated for successive 
three data points in the light curve for 2 $-$ 18 keV energy range
during the dip and non-dip regions. There is a sharp decrease in 
the rms value at the beginning of the
dip which recovers gradually as the flux increases.  

The hardness ratio is computed for all the PPC observations.
It is found that during the dips, the value of the 
hardness ratio is less in comparison to the non-dip regions for all the
observation which indicates that the spectrum of the source is soft during 
the dips. There is no significant change in the value of hardness ratio 
averaged over each orbit. The timing behavior of the source was studied 
by taking the Fourier transform of 1 s and 0.1 s resolution 
mode data. From the power density spectrum (PDS) constructed from
0.1 s time mode data of June 17,  narrow quasi-periodic oscillations 
(QPOs) are detected at 4.5 Hz as shown in Figure 5 along with the 1s  
bin light curve. From the figure, it is clear that the QPOs  
are present in the source when there are no dips in the light 
curve. As the time 
resolution is limited to 1 s for most of the observations, we could 
not study in detail the QPOs in the frequency range of 1 $-$ 10 Hz during 
those periods when the dips were present.

In Table 2, we summerise various properties of the source during the
dip and non-dip periods. The average count rate during the dips is about
half or one-third of the count rate during the non-dip regions. The values of
the hardness ratio and rms variations of the source are always smaller
for the dip periods compared to the non-dip periods.

All the above properties of the source during the non-dip and dip regions
are also seen in the publicly available RXTE/PCA data on 1999 June 08. 
The RXTE/PCA public archive contains only one observation during the radio
flare (from June 8 to June 13) and we have analyzed this data to
substantiate the properties of the dips. The first dip in X-ray band
detected with the PPCs (shown in Figure 2, second panel)
was also seen with RXTE/PCA at the same time. Figure 6 shows the X-ray light 
curve of the RXTE/PCA data in the energy range of 2 $-$ 13 keV 
(upper panel) and the hardness ratio (count rate in 5 $-$ 13 keV 
energy range / count rate in 2 $-$ 5 keV energy range) (lower panel) 
for the source. The PDS for the source for the different regions i.e. regions
during the dips and during the absence of dips is shown in Figure 7. The
first, third and fifth panels in the figure represent the PDS during the
absence of the dips and second and fourth show the PDS during the dips.
A narrow QPO peak at a centroid frequency in the range 4 $-$ 6 Hz is seen in the
non-dip regions which is absent in the dips. The spectrum during the dips
is softer in comparison to the non-dip regions. These properties
are consistent with the PPC observations. 
  
The dips detected in the X-ray light curves with the PPCs are similar to
the dips which occurred after the spike in the X-ray light curve, observed 
on 1997 May 15 and Sept 09 (Mirabel et al. 1998; Markwardt et al. 1999) 
which is described as the ``quiet'' state by Markwardt et al. (1999). 
The properties of the source during this state are similar to those 
found during the dips observed in 1999 June with the PPCs. 

During the multi-wavelength observation of the source carried out in 1997
May 15 and Sept 09, it was found that an X-ray event is followed by 
non-thermal infrared and radio flares. Here we estimate whether a series 
of mini-jets similar to those observed in 1997 May 15 (Mirabel et al. 1998)
associated with the dips can account for the huge radio flare.
From the simultaneous observation of the source in X-ray, IR and radio, 
Mirabel et al. (1998) estimated the value of spectral index of the 
relativistic electrons by applying the van der Laan (1966) model
and derived a relation for the  observed maximum flux density (S$_\lambda$)
 at a given wavelength \(\lambda\) and the time since 
the ejection to reach the maximum flux at a particular wavelength 
(t$_\lambda$) given by
\begin{equation}
S_\lambda~\propto~\lambda^{-(7p+3)/(4p+6)}
\end{equation}
\begin{equation}
t_\lambda~\propto~\lambda^{(p+4)/(4p+6)}
\end{equation}
They found the value of spectral index $p \simeq 0$ by using the observed 
maximum flux density at 3.6 cm and 6 cm during the burst observed on 1997 
May 15. They determined the time for
$t_{6cm} \simeq 0.9$ h and $t_{3.6cm} \simeq 0.65$ h. Using these parameters
we estimate the peak flux at 2.25 GHz ($\lambda$ = 13.3 cm) as
26.5 mJy and t$_{13.6cm}$ as 1.6 hr.
Assuming the time profile of the radio emission as seen by Mirabel
et al. (1998) we estimate the total radio emission at 2.25 GHz 
(\(\lambda\) = 13.3 cm) 
in the radio mini-flare corresponding to a dip in X-ray flux as 
\begin{equation}
F_{13.3}~\simeq~  40 mJy hour.
\end{equation}
The total radio emission at 2.25 GHz during the 1999 June flaring is
estimated by integrating the radio profile (for 6 days)
 by fitting an exponential
to the decay phase of the light curve. If the total radio emission
in the flare observed in 1999 June is a superposition of such mini-flares,
we can calculate the approximate 
number of dips in the X-ray light curve required to produce the radio 
flare as 720. If mini-flares in the radio are associated with the dips 
in the X-ray, this gives a rate of one dip in 12 minutes (in PPC).

\section{DISCUSSION}

 Galactic superluminal sources are ideal astrophysical laboratories
to probe in detail the connection between jets and accretion disks,
ubiquitously thought to be present in Quasars. There were several attempts
in the past to isolate characteristics in the X-ray emission of these sources
(assumed to be coming from an accretion disk) and relate them to the onset of
jet emission. Also, evidence for a sudden mass ejection event from the 
accretion disk is sought in the X-ray emission characteristics. 
Belloni et al. (1997) have
discovered a series of outbursts in GRS 1915$+$105 which were attributed
to ``inner-disk'' evacuation.  They found that the source makes transition
to two intensity levels lasting from a few tens to a few hundred 
seconds with distinct and different inner disk radii
(as obtained from spectral measurements) and they attributed this change
to the disappearance of inner disk due to thermal-viscous instabilities.
Fender et al. (1999) made a detailed analysis of the super-luminal jet ejection
events observed in GRS 1915$+$105 in 1997 October/November and detected
continuous short period (20$-$40 minute) radio oscillations shortly after the
start of the jet emission. They proposed that these are indications of 
repeated ejection of inner accretion disk, quite similar to the events
seen in X-rays by Belloni et al. (1997). A causal connection between disk
and jet was thus attempted.

 Paul et al. (1998) and Yadav et al. (1999) have made a detailed study of
such intensity variations using the data obtained with the IXAE and 
contemporaneous
to the burst events reported by Belloni et al. (1997). They found that
the source spectrum i.e. the ratio between the count rate in 6 $-$ 18 keV 
energy range/count rate in 2 $-$ 6 keV energy range, was softer during
the burst and harder during the quiescent phase.  Yadav et al.
(1999) concluded that the repeated intensity variation cannot be attributed
to inner disk evacuation. This is due to the viscous time scale arguments
as well as due to the fact that the two intensity states are quite similar to
the low-hard and high-soft state of the source. 
They invoked the two component accretion flow model (TCAF) of Chakrabarti
\& Titarchuk (1995) to conclude that the rapid changes are due to the
appearance and disappearance of advective disk covering the standard thin disk
without any requirements of mass ejection. 
Further, these types of hard dips (with a transition time of a few seconds)
are seen almost for about a month in 1997 June (Yadav et al. 1999) when
the radio emission was low. We estimate an average radio emission for
this month as 8 mJy at 2.25 GHz using the GBI data. There was also
no evidence for any flares (flux $<$ 20 mJy at all times).
Hence the causal relationship
between the disk instabilities and jet emission can be treated as not
completely established. 

In the next sub-section we summaries the available X-ray and radio
observations which indicate the disk-jet connections.

\subsection{Radio and X-ray emission in GRS~1915$+$105}

1. There are periods of long durations when both the radio
as well as the X-ray emissions are low. One example is from
1997 January-March when the X-ray flux was $\sim$0.25 Crab and 
2.25 GHz radio flux was 10 mJy. These are classified as 
the radio-quiet hard state observations in Muno et al. (1999).

2. There are durations when the
X-ray and  radio flux  are higher, which are classified as 
radio-loud hard-steady state. They are also known to
exhibit optically thick radio emission and referred to as
the ``plateau'' state (1996 July-August; 1997 October).
For the 1997 October ``plateau'' state, the X-ray flux is
0.50 Crab and 2.25 GHz radio flux is 50 mJy. There is
evidence for an AU scale radio jet observation in this state
(Dhawan et al. 2000).

3. The X-ray flux changes from the steady-hard state to
a flaring-state at various time-scales from a few hours (as seen
in the present work)  to a few months. During this time a variety
of X-ray variations are seen. 

3.1 The X-ray flux teeters between two intensity states
 in a short time (a few second) with a periodicity of 20 $-$ 150 s. 
These are called ``inner-disk oscillations'' by 
Belloni et al. (1997) and ``irregular and quasi-regular'' bursts
by Yadav et al. (1999). There is no evidence for enhanced radio emission
during these events.

3.2 A peculiar morphology of X-ray emission is associated with
radio and infra-red flares. The X-ray emission changes from a 
high oscillating state (at a  period of 10$-$20 s) to a low
hard state (in a time scale of about 100 s). The X-ray intensity 
in the low-hard state gradually increases.  There is a sudden dip
characterized by low intensity, low hardness ratio
 and disappearance of the 0.5 $-$ 10
Hz QPO. The source gradually returns to the high oscillating
state. These 
 are associated with the synchrotron flares in radio (Mirabel
et al. 1998; Fender \& Pooley 1998) and infrared (Eikenberry et al. 1998).
The peak intensities of these flares are in the range $\sim$ 100 $-$
200 mJy from infrared to radio bands (Eikenberry et al. 2000). 
Eikenberry et al. (1998) strongly argue that the onset of
radio/infrared flare is associated with the soft dip rather than
the gradual change to the low-hard state.

3.3 Eikenberry et al. (2000) identify a series X-ray dips
coincident with  faint infrared flares.  
The peak amplitude of the observed mini-infrared 
flares are found to be $\sim$ 0.5 mJy for a duration of a few hundred seconds. 
The period of the soft X-ray dips, producing the observed flares is 
found to be $\sim$ 20 s. These soft dips have the X-ray characteristics
identical to the dips seen during the synchrotron infrared flares
(3.2 above).

4. The huge radio-flares producing super-luminal blobs are associated
with chaotic X-ray variability (as measured by low time resolution
X-ray data). The radio spectrum is steep during these flares. There
is, as yet, no strong morphological identification with detailed
X-ray emission  characteristics.

In the following sections we argue that these soft X-ray dips are
responsible for the superluminal radio flares.

\subsection{Radio flare as a collection of X-ray dips}

 The peculiar dips presented in this paper provide an additional feature
in the X-ray emission which, we argue, is related to mass ejection and the
consequent jet production. There is a vast difference in the nature of the 
X-ray light curves of the source between the 1999 June observation
reported in the present work and those seen in  1997
using PPCs (Paul et al. 1998; Yadav et al. 1999) and RXTE (Belloni et al. 
1997).  During the present observations 
the spectrum was softer as the hardness ratio of the source 
was less, during the dips in comparison to the non-dip regions. 
The observed properties of the source during the non-dip periods i.e.
rms variability in X-ray flux, presence of QPO at a centroid frequency
of 4 $-$ 6 Hz disappear during the dip. 
 There is a gradual return to the accretion-disk properties: the hardness
ratio and the variability characteristics slowly change back to the
pre-dip values (see Figure 4). 

 Morphologically, these dips have properties very similar to those
seen during the dips responsible for the infrared flares 
(Eikenberry et al. 1998; 2000; Mirabel et al. 1998). Fender et al. (1999)
have worked back the onset time of the super-luminal blobs and
during these times the radio emission shows oscillations in a 
time scale of 20 $-$ 30 minutes. Similar radio oscillations
(at similar periods, but at lower intensity) are observed to be 
accompanied by a series of soft X-ray dips (see Fig 10. of Dhawan et
al. 2000). Further, one of the superluminal blobs was thought to originate
from the core on MJD 50750.5 (which is accompanied by radio
oscillations) and the X-ray emission observed on MJD 50751.7 shows
soft X-ray dips (see Muno et al. Fig 1e).

In view of strong evidence for association of such dip events to 
radio emission, it is suggested
that a series of dips can produce the complete radio flare 
lasting for a few days. The onset of the first
dip event detected with the PPCs on June 8, 15.51 UT and independently
with the RXTE coincided with the onset 
of the radio flare within a few hours. It was shown in the
previous section that a superposition of several
disk evacuation events can produce the radio flare if one assumes 
a scaling for energy from single dip events.
It would be interesting to see whether all radio flares are
necessarily accompanied by such X-ray dips interpreted as 
disk evacuation events.

  To produce the observed radio light curve, one needs to assume
that the number of dips produced as a function of time also follows
a similar time profile. Since the IXAE data are obtained for only
5 of the 14 orbits everyday, observations are not continuous
enough to conclusively establish this hypothesis. It may be pointed out
that during
the superluminal jet events of 1997 October $-$ November, the frequency
of radio oscillations decreased from 2.9 hr$^{-1}$ on MJD 50750.5 (when
the radio flux was 200 mJy) to 1.9 hr$^{-1}$ on MJD 50752.5 (when the
radio flux decreased to 120 mJy - see Fig. 7 of Fender et al. 1998). The
smooth radio light curve and the steep spectrum could be due to the 
movement of the ejecta in the interstellar medium. The radio flares of 1997
October $-$ November  seen by Fender et al. (1998) is the superposition 
of at least four
ejecta and the start of each ejecta is associated with a series of radio
oscillations.

The mass and energy estimates of the superluminal blobs
emitted by GRS 1915+105 (Rodriguez \& Mirabel 1999) is too large
to be caused by a single isolated event in the accretion disk, and
a series of accretion disk driven events would be required, as
suggested by Fender et al. (1999). The series of dips that we
have observed could provide the necessary energy for the superluminal
blobs, if they occur in a rapid series. A continuous X-ray monitoring 
during a radio flare will clarify this question. Also, the dips seen during 
the later part of the radio flare are of shorter durations and these
events may not be ejecting matter in sufficient quantities. It is quite
conceivable that only the long duration dips with a gradual recovery are
responsible for jet emission. The start of the dips always shows
similar observable parameters like count rates and hardness ratio indicating
a causal relationship between disk parameters and the onset of dip events. 

\subsection{TCAF model for the advection accretion}

We attempt to interpret the present results in the light of the
TCAF model for the advective accretion disk around the black holes.

It has been recognized that very near to the black hole, accretion is
necessarily advective and far away from the compact object it is beleived 
that the accretion is through a geometrically thin accretion disk. In the
ADAF (Advection Dominated Accretion Flow) model of Narayan \& Yi (1994)
the changeover occurs at a transition radius (r$_{tr}$) whereas in the Two
Component Accretion Flow (TCAF) model of Chakrabarti \& Titarchuk (1995)
a standing shock wave or a centrifugal barrier dominated dense region at
R$_0$ separates these two. Das \& Chakrabarti (1999) have included the
effect of outflow in the TCAF model. GRS 1915$+$105 changes from hard
to soft states at a variety of time scales and the observed properties
of the source can be interpreted as the variation in R$_0$.

\subsubsection{Time scales of the dip events}

Yadav et al. (1999) derived various time 
scales for the quiescent period, decay time and rise time for the different 
types of bursts observed by equating the inner disk radius(R$_{in}$) to 
R$_{o}$ from where the advection dominated halo component covers the thin 
accretion disk during the quiescent state. They obtained the viscous time 
scale for the standard \(\alpha\) disk (t$^{d}_{vis}$) as

\begin{equation}
 t^{d}_{vis} = 4.3 \times 10{^{-4}} \alpha^{-1} \dot{m}^{-1}_{d}m^{-1}R^{2}_{o}
\end{equation}
where \.{m}${_d}$ is in the unit of Eddington accretion rate, m is the
mass of the source in the unit of solar mass and R$_{o}$ in km. They
calculated this time scale (t$^{d}_{vis}$) to be of the order a few hundred seconds 
for \.{m}${_d}$ = 1, m = 10, \(\alpha\) = 0.01 and R = 300 km. 

Similarly, the time scale for the advection disk for halo component
t$^{h}_{vis}$ (Yadav et al. 1999) as

\begin{equation}
t^{h}_{vis} = 4.9 \times 10{^{-6}} \alpha^{-1} m^{-1/2}R^{3/2}_{o}
\end{equation}

Using the parameters used to derive t$^{d}_{vis}$
the free fall time scale is calculated to be order of seconds, which agrees
with the decay time of the observed dips.

The recovery time of the dips ($\sim$ 110 s) agrees with the calculated
viscous time scale of the standard disk t$^{d}_{vis}$.
 
The time-scale of the dip onset 
(a few seconds) is comparable to the free-fall time scale as well as the 
advective disk time scale (equation 5), for typical advection dominated disk 
parameters.  Hence we can conclude that all 
the accretion disk characteristics suddenly disappear during the dips 
and they reappear in a gradual way. We feel that a sudden disk evacuation 
and a gradual refilling is a natural explanation for this observation.

\subsubsection{TCAF explanation for the outflow}

 To produce radio jets, a large amount of matter has to be expelled from the
disk and this has to be accelerated to relativistic velocities. 
Das \& Chakrabarti (1999) have proposed a combined 
inflow/ outflow model in which the computation is done using combinations 
of exact transonic inflow and outflow solutions. 
assuming free-falling conical polytropic inflow and isothermal outflows,
they estimated the ratio of out flowing and in-flowing rate to be
\begin{equation}
\frac{\dot{M}_{\odot}}{\dot{M}_{in}}= R_{\dot{m}} =~\frac{\Theta_{out}}{\Theta_{in}} ~\frac{R}{4}~~e^{-(f_0 -~\frac{3}{2})}~ f_0^{3/2}
\end{equation}
where \(\Theta\)$_{out}$ and \(\Theta\)$_{in}$ are the solid angles of the
outflow and inflow respectively, R is the compression ratio of the in-flowing
matter which is a function of the flow parameters such as specific energy
and angular momentum (Chakrabarti 1990) and f$_0$ is given by 
\begin{equation}
f_0 = \frac{R^2}{R-1} = \frac{(2n+1)~ R}{2n}
\end{equation}
\textit{n} is the polytropic constant = 1/(\(\gamma\)$-$1),~~ \(\gamma\) being 
the adiabatic index. 

They found that the outflow rate depends on the initial parameters 
of the flow and in some cases the outflow
rate can be higher than the inflow rate leading to disk evacuation. We propose
that the dips that we have observed are a result of such disk 
evacuation phenomenon. It would be interesting to carry out a detailed spectral
and timing analysis to extract the exact disk parameters leading to a
disk evacuation event. 

\subsection{Jet acceleration}

It is possible that the accretion disk magnetic field is responsible for 
accelerating the disk evacuated matter into radio emitting jets.
Meier et al. (1997) have proposed a magnetic switch that can generate
superluminal jet ejection. They have shown that when the coronal Alfven
velocity exceeds a critical value, the jet velocity can become 
relativistic.  We have calculated the Alfven velocity by assuming the 
advection dominated disk parameters given by Narayan et al. (1998).

The expression for Alfven velocity (V$_{A}$) and Escape velocity (V$_{esc}$) are 
\begin{equation}
 V_{A} = \frac {B}{\sqrt{4\pi\rho}} ,
\end{equation}    
\begin{equation}
V_{esc} = \sqrt{\frac {2GM}{r}}= \frac{c}{\sqrt{r}}
\end{equation}    
where \textit{B} is the magnetic field, \textit{r} is the radial distance, 
\(\rho\) is the plasma density and 
\textit{r} is the distance in the unit of Schwarzschild radius.

Assuming an equipartition with the
magnetic energy, the expression for B and n$_e$ (number density) are given by 
(Narayan et al. 1998)
\begin{equation}
B \simeq 7.8 \times 10{^8}~\alpha{^{-1/2}}~m{^{-1/2}}~\dot{m}{^{1/2}}~r{^{-5/4}}~~~ G
\end{equation}    
\begin{equation}
n_e \simeq 6.3 \times 10{^{19}}~\alpha{^{-1}}m{^{-1}}~\dot{m}~r{^{-3/2}}~~cm{^{-3}}
\end{equation}    
where \(\alpha\) is Shakura \& Sunyaev viscosity parameter,
\textit{m} is in the unit of M$_{\odot}$, \\
and \textit{$\dot{m}$} is in the unit of Eddington accretion rate.

For equipartition of magnetic field (i.e. \(\beta\) = 0.5), \\ 
$~~~~~~~~~~~~~~~~$\(\alpha\) = c~( 1$-$\(\beta\))~\(\simeq\) 0.3~~~~~~(for c~\(\sim\)~0.5$-$0.6)

Taking the values of \textit{m} as 10 and \textit{$\dot{m}$} as 0.1 to 1 for GRS 1915$+$105, we estimated the magnetic field and the density (assuming one proton corresponding to one electron in the plasma) as

$~~~~~~~~~~~~~~$B = 4.51 \(\times\) 10{$^8$} ~\(\dot{m}\){$^{1/2}$}~r{$^{-5/4}$}~~ G\\ 
$~~~~~~~~~~~~~~~~~~~~~$\(\rho\) = 3.5 \(\times\) 10{$^{-5}$}~\(\dot{m}\)~r{$^{-3/2}$} g/cc  

Using the values of B and \(\rho\), the Alfven velocity V$_A$ is calculated to be 
\begin{equation}
 V_{A} = \frac {B}{\sqrt{4\pi\rho}} = \frac{2.1~\times~10{^{10}}}{\sqrt{r}}~ \simeq ~  \frac{2}{3}~\frac{c}{\sqrt{r}}
\end{equation}

Now, the ratio between V$_A$ and V$_{esc}$ is 
\begin{equation}
\frac{V_A}{V_{esc}}~=~\frac{2}{3}
\end{equation}

We find that the Alfven velocity is near the critical
velocity for advection dominated flows indicating the operation of the
magnetic switch. It should be noted that just before the 
detection of the superluminal jet ejection in GRS 1915$+$105 in November
1997 by Fender et al. (1998), the source was in a low-hard state as well as a radio-loud state. We can
envisage a scenario where the accretion disk condition during a  low-hard 
state makes the magnetic switch to operate and the ejected material
is accelerated to relativistic  velocities.  These accelerated blobs
can move in the interstellar medium to give the steep spectrum superluminal
ejecta. The transition period from a low-hard to a high-soft state
provides critical conditions for a series of disk evacuation events
to occur so that sufficient matter is ejected to produce relativistic
jets. A continuous monitoring in X-ray and radio bands
during a radio flare will clarify most of these questions.

\section*{Acknowledgments  }

We express sincere thanks to the referee for his useful
comments and suggestions which improved the contents and presentations
of the paper.
We acknowledge the contributions of the scientific and technical staff
of TIFR, ISAC and ISTRAC for the successful fabrication, launch and 
operation of the IXAE. It is a pleasure to acknowledge constant  support 
of Shri K. Thyagarajan, Project Director IRS$-$P3 satellite, Shri J. D. Rao
and his team at ISTRAC,
Shri P. S. Goel, Director ISAC and the Director of the ISTRAC.
We thank the RXTE/ASM and NSF-NRAO-NASA Green Bank Interferometer group 
for making the data publicly available. The Green Bank Interferometer is 
a facility of the National Science Foundation operated by the NRAO in 
support of NASA High Energy Astrophysics programs.

\clearpage

\begin{figure}
\centering
\psfig{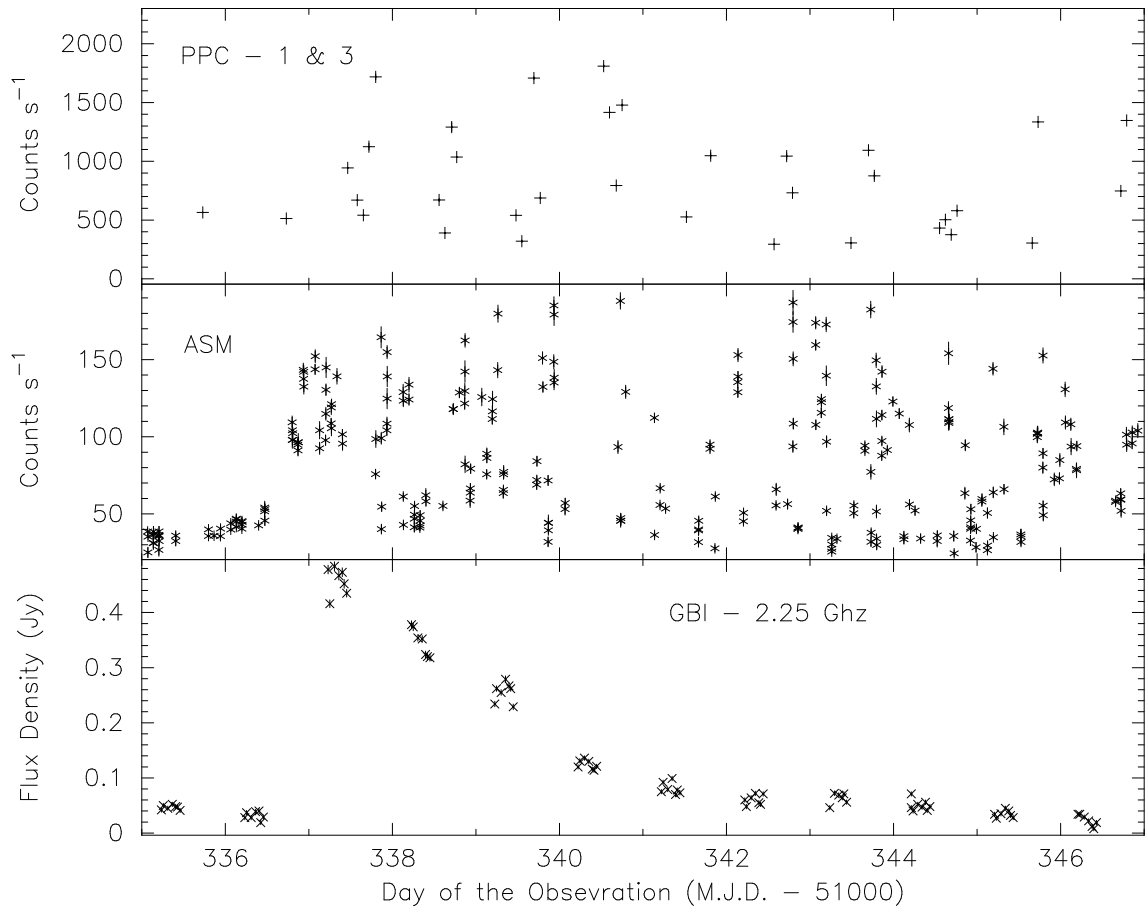}
\caption[fig1.eps]{The X-ray light curve for GRS 1915$+$105 with the 
PPCs (averaged over each orbit) in the energy range 2$-$18 keV, with 
RXTE ASM in the range 1.3$-$12.2 keV and radio flux at 2.25 GHz with 
NSF-NRAO-NASA Green Bank Interferometer, during the observation 
of the source by the PPCs in IXAE in 1999 June.}\label{fig1}

\end{figure}

\begin{figure}
\centering
\psfig{file=f2.ps,width=12cm,angle=-90}
\caption[fig2.eps]{}
\label{fig2a}

\end{figure} 

\begin{figure}
\centering
\psfig{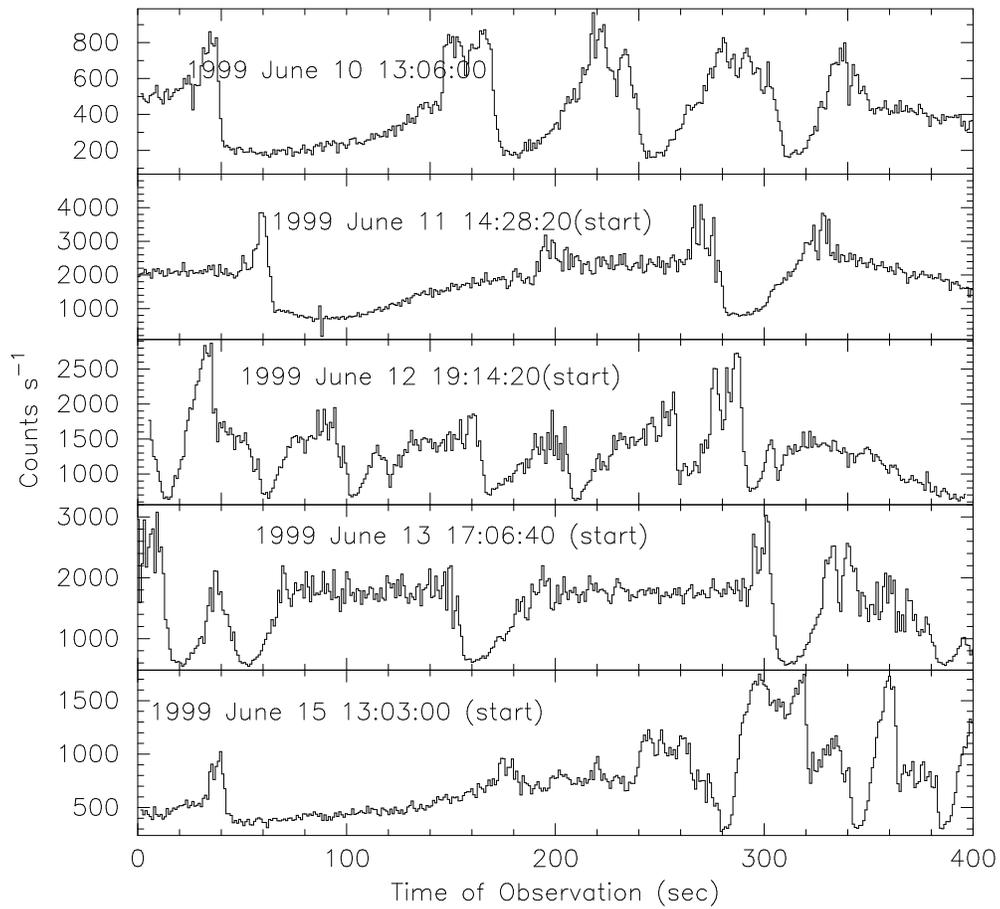}
\caption[fig2.eps]{Fig. 2 and 3 show the light curves of GRS 1915$+$105 
with the PPCs with 1 s bin size in the energy range 2$-$18 keV. The presence 
of different types of dips with different periods in the light curve of the 
source are shown.}\label{fig2b}

\end{figure}

\begin{figure}
\centering
\psfig{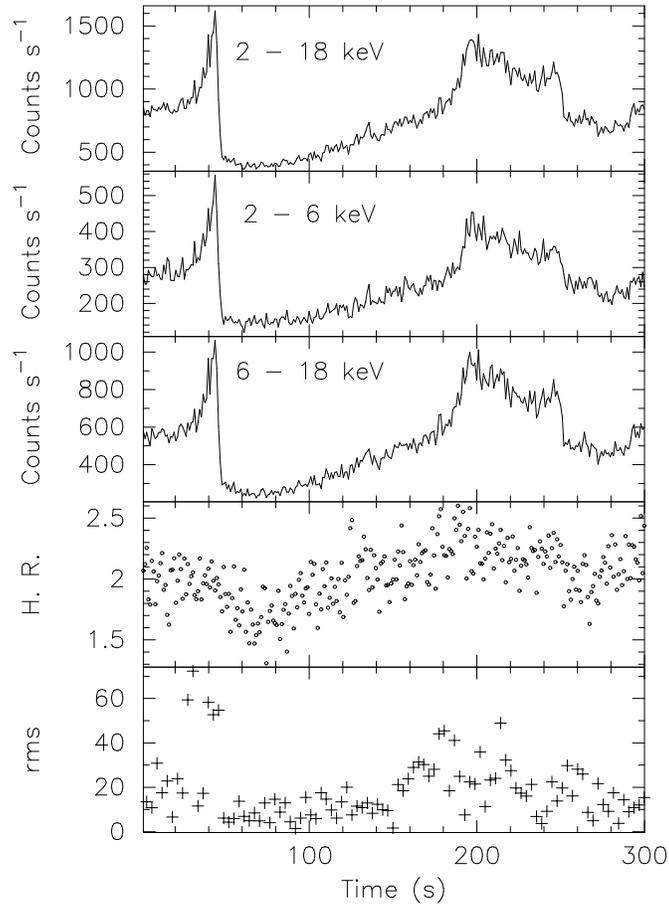}
\caption[fig3.eps]{The light curves of GRS 1915$+$105 during the dips 
observed with the PPCs with 1 s bin size in the energy range 2$-$18 keV, 
2$-$6 keV and 6$-$18 keV. The dip is identical in shape in all the 
energy ranges. The hardness ratio ( H. R.) is plotted in the fourth 
panel of the figure which indicates the soft nature of the spectrum 
during the dips. The lower panel shows the variation in the rms of 
the source during the dip and non-dip regions.}\label{fig3}

\end{figure}

\begin{figure}
\centering
\psfig{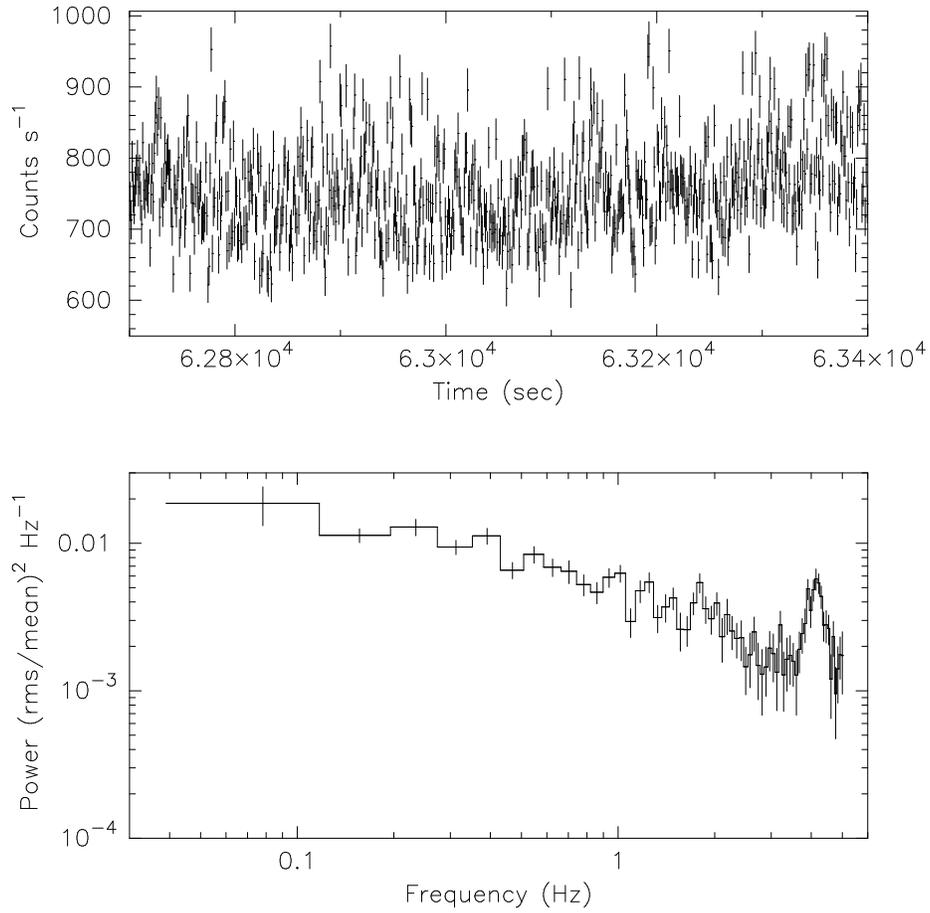}
\caption[fig4.eps]{The light curve and PDS of GRS 1915$+$105 with the PPCs
in the energy range 2$-$18 keV  on 1999 June 17. The presence of a QPO
peak at frequency 4.5 Hz is seen during the non-dip period in the
X-ray light curve. }\label{fig4}

\end{figure}

\begin{figure}
\centering
\psfig{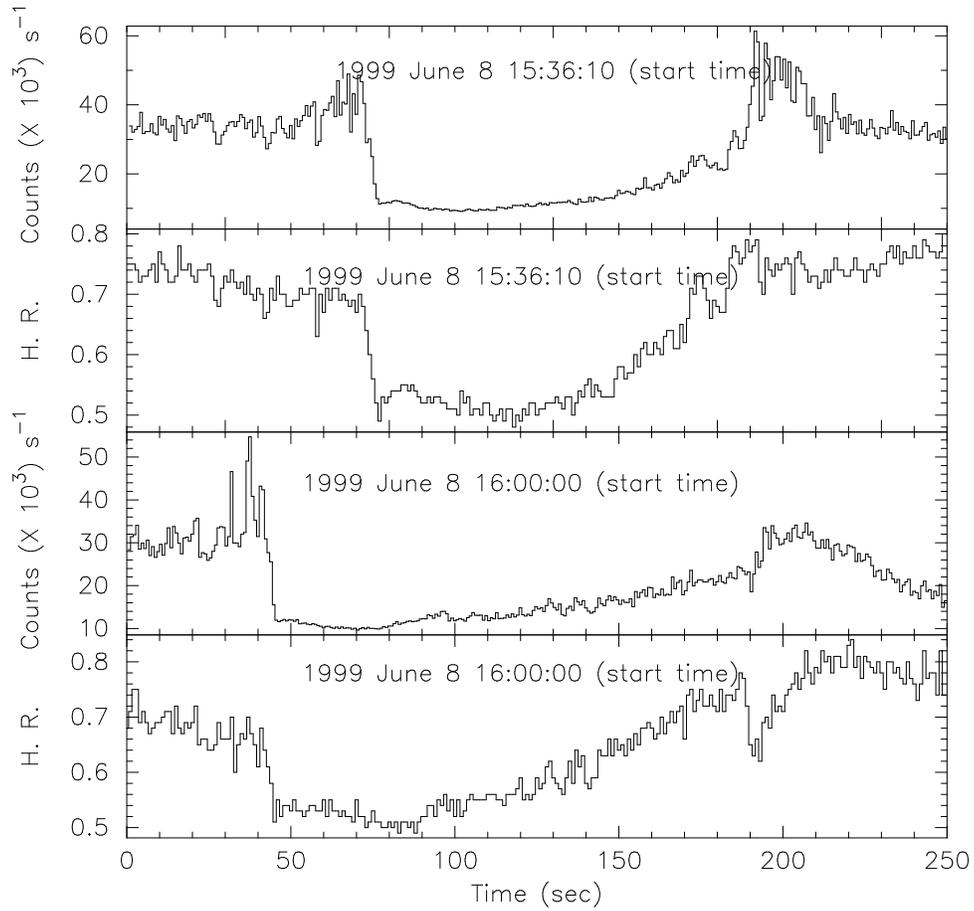}
\caption[fig6.eps]{The light curve of GRS 1915$+$105 obtained with the 
RXTE/PCA data on MJD 51337 in 2 $-$ 13 keV energy range with 0.8 s time 
bin is shown for two dips ( A \& C). The hardness ratio (H. R.) of the 
source i.e. countrate in 5 $-$ 13 keV energy range / countrate in 
2 $-$ 5 keV energy range is also shown (B \& D).}\label{fig6}

\end{figure}

\begin{figure}
\centering
\psfig{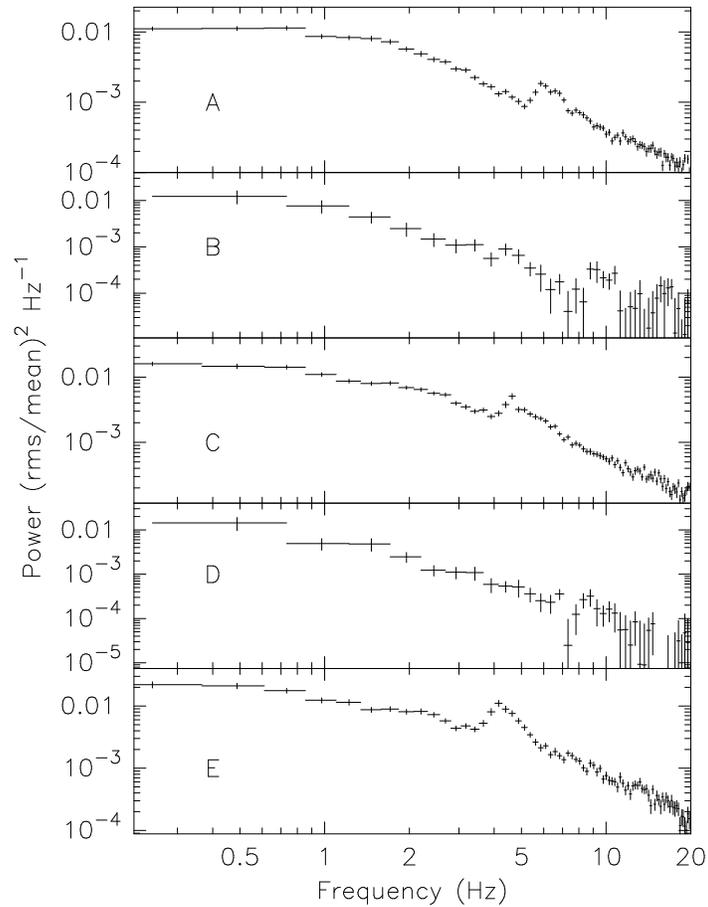}
\caption[fig7.eps]{The PDS of the source GRS 1915$+$105 in the energy range
5 $-$ 13 keV for different regions of the light curve.  The panels A, 
C, and E represent the PDS during the non-dip periods i.e. earlier 
to the first dip, between the two dips and beyond the second dip 
respectivly, as seen with the RXTE/PCA on 1999 June 8. The panels B 
and D represent the PDS during the dip periods in the light curve. 
The figure shows the presence of QPOs in the frequency range 4 $-$ 6 
Hz during the non-dip regions which are absent during the dips.}\label{fig7}

\end{figure}

\clearpage

\begin{deluxetable}{llllccll}
\footnotesize
\tablecaption{Log of X-ray Observation of GRS 1905$+$105 with IXAE\label{tbl-1}}
\tablewidth{0pt}
\tablehead{ 
\colhead{Obs. Date,} & \colhead{Start time} & \colhead{End Time} &\colhead{Count} & \colhead{Source}  & \colhead{Duration} \\ 
\colhead{1999 June}  & \colhead{(UT)} & \colhead{(UT)} & \colhead{rate}  & \colhead{intensity}  & \colhead{of dips (s)}  \\
\colhead{} & \colhead{} & \colhead{} & \colhead{}  & \colhead{(in Crab)} &\colhead{}}
\startdata

06        &17:01   &17:18    &565  &0.686     &---\nl   
07        &17:31   &17:48    &513  &0.623     &--- \nl     
08	  &11:02   &11:15    &944  &1.146     &--- \nl
          &13:45   &14:03    &670  &0.814     &--- \nl 
          &15:28   &15:45    &541  &0.657     &125 \nl
          &17:09   &17:20    &1124 &1.365     &140, 40 \nl 
          &18:54   &19:06    &1718 &2.086     &160 \nl 
09        &13:24   &13:42    &671  &0.814     &145 \nl 
          &15:07   &15:24    &399  &0.485     &--- \nl 
          &16:48   &17:05    &1291 &1.568     &85,60 \nl 
          &18:28   &18:43    &1036 &1.258     &--- \nl 
10        &11:21   &11:39    &540  &0.656     &--- \nl 
          &13:03   &13:21    &320  &0.389     &110,50,40,30 \nl 
          &16:40   &16:44    &1708 &2.074     &120 \nl
          &18:07   &18:24    &688  &0.835     &110 \nl 
11        &12:47   &12:59    &1809 &2.196     &--- \nl 
          &14:23   &14:41    &1416 &1.719     &130,40 \nl 
          &16:06   &16:22    &794  &0.964     &--- \nl 
          &17:46   &18:04    &1478 &1.794     &120 \nl 
12        &12:20   &12:37    &527 &0.640    &10 no.s (20 $-$ 40 s) \nl 
          &19:13   &19:23    &1048 &1.272   &7 no.s (20 $-$ 40 s) \nl 
13        &13:41   &13:59    &295  &0.358  &160,40,30,20 \nl 
          &17:06   &17:23    &1044 &1.267  &40,30,30 \nl 
          &18:49   &19:01    &732  &0.889  &--- \nl 
14        &11:38   &13:55    &306  &0.372 &120,40,20,20 \nl 
          &16:45   &17:01    &1094 &1.328  &100,30,20,40 \nl 
          &18:23   &18:40    &876 &1.063  &30,20,40,30  \nl  
15        &12:59   &13:16    &432  &0.524  &120,20,20,20 \nl 
          &14:41   &14:57    &503  &0.611 &5 no.s (20 $-$ 30 s) \nl 
          &16:27   &16:40    &376  &0.457 &--- \nl 
          &18:01   &18:19    &580  &0.704 &3 no.s (20 $-$ 30 s) \nl 
16        &15:59   &16:21    &305 &0.37  &--- \nl 
          &17:41   &18:00    &1335 &1.62  &--- \nl 
17        &17:18   &17:40    &748  &0.908 &--- \nl 
          &18:49   &19:00    &1347 &1.635 &--- \nl 
\tablenotetext{}{ }
\enddata
\end{deluxetable}

\clearpage

\begin{deluxetable}{lcccccccc}
\footnotesize
\tablecaption{{}}
\tablewidth{0pt}
\tablehead{ 
\colhead{Obs. Date,} & \colhead{Count rate} & \colhead{Count rate} &\colhead{~rms} & \colhead{~rms}  & \colhead{Avg H.R.} & \colhead{Avg H.R.}\\ 
\colhead{1999 June}  & \colhead{during dip} & \colhead{during non-dip} & \colhead{during dip}  & \colhead{during non-dip}  & \colhead{during dip}  & \colhead{during non-dip}}
\startdata
08 15:36     &253.2   &713.5   &13.9   &28.6    &---   &--- \nl
~~~~17:09    &781.5   &1540.6  &2.54   &8.69    &---   &---\nl
~~~~18:54    &854.2   &1842    &2.5    &5.95    &---   &---\nl
09 13:24     &361.3   &791.6   &2.32   &3.27    &1.8   &2.1\nl
~~~~16:48    &782.2   &2214.8  &2.58   &5.26    &1.63  &1.95\nl
10 13:03     &198.1   &555.77  &1.38   &4.14    &1.8   &2.13 \nl
~~~~16:39    &409.3   &814.1   &1.14   &2.10    &1.75  &1.92 \nl
~~~~18:09    &387.8   &816.3   &1.26   &3.17    &1.67  &1.865 \nl 
11 14:23     &725.3   &2029.2  &1.97   &4.52    &---   &---\nl
~~~~17:46    &257.1   &615.0   &1.91   &1.77    &---   &---\nl
12 12:20     &267.2   &635.8   &2.6    &2.36    &1.9   &2.1\nl
~~~~19:13    &724.3   &1450.8  &1.78   &3.89    &1.6   &1.98\nl
13 13:41     &177.9   &235.9   &1.81   &2.32    &1.95  &1.98\nl
~~~~17:06    &691.6   &1769.9  &2.75   &3.98    &1.75  &2.04\nl
14 11:39     &221.4   &526.6   &2.2    &4.45    &1.98  &2.52\nl
~~~16:44     &721.2   &1434    &1.95   &4.79    &1.74  &2.18\nl
~~~~18:23    &650.5   &1563.2  &2.64   &5.8     &1.75  &2.06\nl
15 12:59     &400.5   &784.7   &1.79   &2.42    &1.89  &2.40\nl
~~~~14:41    &286.3   &462.4   &1.11   &2.07    &1.65  &1.95\nl
\tablenotetext{}{ {\bf Note:} H. R. is for Hardness Ratio.}
\enddata
\end{deluxetable}


\begin{thebibliography}{}

\bibitem[Agrawal et al. 1996]{}
Agrawal, P. C., et al. 1996, IAU Circ. 6488 

\bibitem[Agrawal (1998)]{1}
Agrawal, P. C., {\it Perspective in HEAA, Proceedings of the International Colloquium}, Aug. 12-17, 1998, TIFR, Mumbai


\bibitem[Belloni et al. (1997)]{bell:97}
Belloni, T., Mendez, M., King, A. R., van der Klis, M., van Paradijs, J. 1997, ApJ, 479, L145

\bibitem[Castro-Tirado et al. 1992]{cast:92}
Castro-Tirado, A.J., Brandt, S., \& Lund, N. 1992, IAU Circ., 5590.

\bibitem[]{} 
Chakrabarti, S. K., 1990, Theory of Transonic Astrophysical Flows (Singapore: World Sci.)

\bibitem[]{} 
Chakrabarti, S. K. \& Titarchuk, Lev G. 1995, ApJ, 455, 623


\bibitem[Chen et al. 1997]{chen:97}
Chen, X., Swank, J. H., Taam, R. E.  1997, ApJ, 477, L41

\bibitem[]{}
Das, T. K. \& Chakrabarti, S. K. 1999, Class. Quantum Grav., 16, 3879

\bibitem[]{}
Dhawan, V., Mirabel, I. F., \& Rodriguez, L. F. 2000, ApJ submitted

\bibitem[Eikenberry  et al. 1998]{eike:98}
Eikenberry, S. S., Matthews, K., Morgan, E. H., Remillard, R. A. \& Nelson, 
R.W.  1998, ApJ, 494, L61

\bibitem[Eikenberry  et al. 2000]{}
Eikenberry, S. S., Matthews, K., Muno, M., Blanco, P. R., Morgan, E. H., \& Remillard, R. A. 2000, astro-ph/0001472  

\bibitem[]{}
Fender, R. P., Garrington, S. T., McKay, D. J., Muxlow, T. W. B., Pooley, G. G., Spencer, R. E., Stirling, A. M., Waltman, E. B.  1999, MNRAS, 304, 865

\bibitem[]{}
Fender, R. P., \& Pooley, G. G. 1998, MNRAS, 300,573

\bibitem[]{}
Feroci, M., Matt, G., Pooley, G. et al. 1999, A\&A, 351, 985

\bibitem[Harmon et al. 1997]{harm:97}
Harmon, B. A., Deal, K. J., Paciesas, W. S., Zhang, S. N., Robinson, C. R., Gerard, E., Rodriguez, L. F. \& Mirabel, I. F.  1997, ApJ, 477, L85

\bibitem[]{}
Markwardt, C. B., Swank, J. H. \& Taam, R. E. 1999, ApJ, 513, L37

\bibitem[]{}
Meier, D. L., Edgington, S., Godon, P., Payne, D. G., \& Lind, K. R. 1997, Nat, 388, 350

\bibitem[Mirabel \& Rodriguez 1994]{mira:94}
Mirabel, I. F. \& Rodriguez, L. F. 1994, Nat, 371, 46

\bibitem[]{}
Mirabel, I. F., Dhawan, V., Chaty, S. et al. 1998, A\&A, 330, L9

\bibitem[Morgan \& Remillard  1996]{morg:96}
Morgan, E. H. \& Ramillard, R. A. 1996, IAU Circ., 6392


\bibitem[Muno et al. 1999]{muno:97}
Muno, M. P., Morgan, E. H., \&  Ramillard, R. A. 1999, ApJ, 527, 321 

\bibitem[]{}
Narayan, R., Mahadevan, R. \& Quataert, E. 1998, The Theory of Black Hole
Accretion Discs, eds. M. A. Abramowicz, G. Bjornsson \& J. E. Pringle

\bibitem[]{}
Narayan, R., \& Yi, I. 1994, ApJ, 428, L13


\bibitem[Paul et al. 1998]{paul:98a}
Paul, B., Agrawal, P. C., Rao, A. R. et al. 1998, ApJ, 492, L63


\bibitem[Rao  et al. 1998]{rao:98}
Rao, A. R., Agrawal, P. C., Paul, B. et al. 1998, A\&A, 330, 181


\bibitem[Rao  et al. 1998]{rao:98}
Rodriguez, L. F. \& Mirabel, I. F.  1999, ApJ, 511, 398

\bibitem[Trudolyubov  et al. 1999]{trud:98}
Trudolyubov, S., Churazov, E., \& Gilfanov, M. 1999, Astr. L., 25, 718 

\bibitem[]{}
van der Laan, H., 1966, Nat, 211, 1131

\bibitem[Yadav et al. 1999]{yadav:99}
Yadav, J. S., Rao, A. R., Agrawal, P. C., Paul, B., Seetha, S., \&
Kasturirangan, K. 1999, ApJ, 517, 935

\end{thebibliography}
\end{document}